\documentclass[12pt]{iopart}
\usepackage[pdftex]{graphicx}

\usepackage[active]{srcltx}
\usepackage{subfigure}
\usepackage{pifont}

\newcommand{\remark}[1]{}

\newcommand{\Ca}{$^{40}$Ca$^+\,$}

\usepackage{iopams}
\begin{document}

\title[Ion-trap heating rates]{Fabrication and heating rate study of microscopic surface electrode ion traps}

\author{N.~Daniilidis$^{1,2}$, S.~Narayanan$^{1,2}$, S.~A.~M\"oller$^{1,2}$, R.~Clark$^{2,3}$, T.~E.~Lee$^{4}$, P.~J.~Leek$^{5}$, A.~Wallraff$^{5}$, St.~Schulz$^{3}$, F.~Schmidt-Kaler$^{6}$, and H.~H\"affner$^{1,7}$}

\address{$^{1}$ Dept. of Physics, University of California, Berkeley, CA 94720, USA\\}
\address{$^{2}$ Institut f\"ur Quantenoptik and Quanteninformation, Innsbruck, Austria\\}
\address{$^{3}$ Center for Ultracold Atoms, Massachusetts Institute of Technology, Cambridge, MA, USA\\}
\address{$^{4}$ Department of Physics, California Institute of Technology, Pasadena, CA 91125}
\address{$^{5}$ Department of Physics, ETH Z\"urich, CH-8093 Z\"urich, Switzerland\\}
\address{$^{6}$ Universit\"at Mainz, QUANTUM, D-55128 Mainz, Germany\\}
\address{$^{7}$ Materials Sciences Division, Lawrence Berkeley National Laboratory, Berkeley, CA 94720, USA\\}
\ead{hhaeffner@berkeley.edu}

\begin{abstract}
We report heating rate measurements in a microfabricated gold-on-sapphire surface electrode ion trap with trapping height of approximately 240~$\mu$m. Using the Doppler recooling method, we characterize the trap heating rates over an extended region of the trap. The noise spectral density of the trap falls in the range of noise spectra reported in ion traps at room temperature. We find that during the first months of operation the heating rates increase by approximately one order of magnitude. The increase in heating rates is largest in the ion loading region of the trap, providing a strong hint that surface contamination plays a major role for excessive heating rates. We discuss data found in the literature and possible relation of anomalous heating to sources of noise and dissipation in other systems, namely impurity atoms adsorbed on metal surfaces and amorphous dielectrics.
\end{abstract}

\section{Introduction}

Trapped ions provide a promising candidate to pursue quantum information processing. Successful implementations of basic quantum protocols as well as the creation of entangled states support this view \cite{Haeffner2008,Blatt2008}.  A promising route to scalability of ion-trap based quantum information processing was proposed in which segmented trap electrodes allow splitting, shuttling, and recombination of ion crystals \cite{Kielpinski2002}. To overcome the difficulties of assembling three-dimensional trap structures and simplify fabrication, several groups are developing planar (surface) ion traps where all electrodes lie within one plane and can be constructed using standard microfabrication methods \cite{Chiaverini2005,Seidelin2006,Britton2006,Pearson2006,Labaziewicz2008,Leibrandt2009,Allcock2009}. In particular, recently the NIST group successfully transported ions through junctions in such planar traps \cite{Amini2009}.

Despite the recent progress, there are still a number of unresolved difficulties with planar traps. In the effort to develop this technology, one strives for miniaturization in order to easily achieve the high trap frequencies required for fast splitting of ion crystals \cite{Chiaverini2005}, and to increase coupling rates in approaches where trapped ions are coupled to \cite{Tian2004} or via solid state elements \cite{Daniilidis2009}. Thus, planar traps with ion-surface separations of less than 100~$\mu$m are pursued. However, the observed ion heating rates are more than 3 orders of magnitude higher than those expected from Johnson noise considerations \cite{Monroe1995,Wineland1998}, imposing a  major obstacle in developing this approach.

This so-called ``anomalous'' heating in ion traps is usually discussed in terms of the ``patch potential'' model \cite{Turchette2000}. The essential features of this model are that the electric field noise responsible for heating of the ion motion  arises from a large number of randomly fluctuating sources and that the electric field of individual sources scales with source-ion separation as an electrical dipole field.  These two features reproduce the observed $1/d^4$ scaling of the electric field noise with the separation,  $d$, between the ion and trap electrodes. Besides the scaling with distance, the noise spectral density of anomalous heating is believed to scale as $1/f$ with frequency \cite{Turchette2000} and experiments indicated increase of noise on particular traps with time, possibly due to contamination of the trap electrodes \cite{Turchette2000,DeVoe2002}. More recently it has been found that the noise is drastically reduced at low temperatures \cite{Labaziewicz2008}, and seems to be related to thermally activated processes \cite{Labaziewicz2008a}, further supporting the $1/f$ nature.

This phenomenology suggests a connection with noise in various other systems. Electric field noise is a factor in nanomechanics \cite{Li2007}, single spin detection \cite{Mamin2003}, and measurement of weak forces (e.g. gravitation) \cite{Speake2003}.  Moreover charge noise limits the performance of nanoelectronic and quantum electronic devices, such as single electron transistors \cite{Zimmerli1992}, Josephson qubits \cite{Astafiev2004}, superconducting coplanar resonators \cite{Gao2008,Kumar2008, OConnell2008}, and quantum dots \cite{Hayashi2003,Gorman2005}. In these systems decoherence is believed to be caused by tunneling two level systems (TLS) which possess a dipole moment \cite{Shnirman2005,Martinis2005}. Such TLS are common in amorphous dielectrics and are known to contribute to  dielectric losses and attenuation of phonons \cite{Anderson1972,Phillips1972, Phillips1987}. The latter mechanism has also been suggested to be relevant in dissipation encountered in efforts to laser cool dielectric mechanical resonators \cite{Arcizet2009}.

In a different direction, in measurements of non-contact friction using metalized atomic force microscope cantilevers close to metal surfaces, dissipation 9 to 11 orders of magnitude higher than expected for clean metal surfaces is observed \cite{Dorofeyev1999, Persson2000,Volokitin2005}, albeit in a different frequency and distance regime than those accessible with ion traps. This effect is suggested to be due to impurity atoms adsorbed on the metal surfaces, and it is suggested that less than one atomic monolayer of adsorbate is sufficient to produce this effect \cite{Volokitin2005}. In addition, the initially high level of dissipation close to metal surfaces has been found to further increase by up to two orders of magnitude upon deposition of dielectrics on a gold surface\cite{Kuehn2006}.

Finally, magnetic traps for neutral atoms also suffer from electromagnetic field noise \cite{Jones2003}. In this case, magnetic fields in the radio or microwave frequency range couple to the atomic spins and result in atom loss by causing spin flips. Contrary to the situations discussed above, this mechanism is well understood and accounted for by Johnson noise on the metallic trap electrodes \cite{Henkel1999}.

Here we study anomalous heating in ion traps. We describe a simple method for fabrication of planar traps and measure  their heating rates using the Doppler recooling method \cite{Wesenberg2007}. Taking advantage of the segmented trap geometry,  we measure heating rates at various positions above the trap and find strong time and position dependence, which indicates  a major role of surface contamination in excess heating rates. 
The heating rates in the ion-loading region of the trap have increased by more than one order of magnitude over the time span of a few months. In contrast, the heating rates sufficiently far from the loading region remain low throughout our study. 

Finally, we present a phenomenological model for anomalous heating in ion-traps that matches with the above general context. This is done by adapting the patch potential model assuming a distribution of oscillating microscopic dipoles on the trap electrodes, with properties such that the $1/f$  nature of noise in ion traps is reproduced. Using this model we estimate the density of electrical dipole sources on a trap electrode surface that would give rise to heating rates reported in the literature. In this context, we discuss the possible relevance of impurity atoms adsorbed on trap electrodes and tunneling TLS to anomalous heating in ion traps. Our model also explains why no ``anomalous'' heating effects are observed in magnetic atom traps.

\section{Experimental Setup}
\label{sec:setup}

\begin{figure}[!tb]
\begin{center}
\subfigure{(a)\;\includegraphics[width=0.4\textwidth]{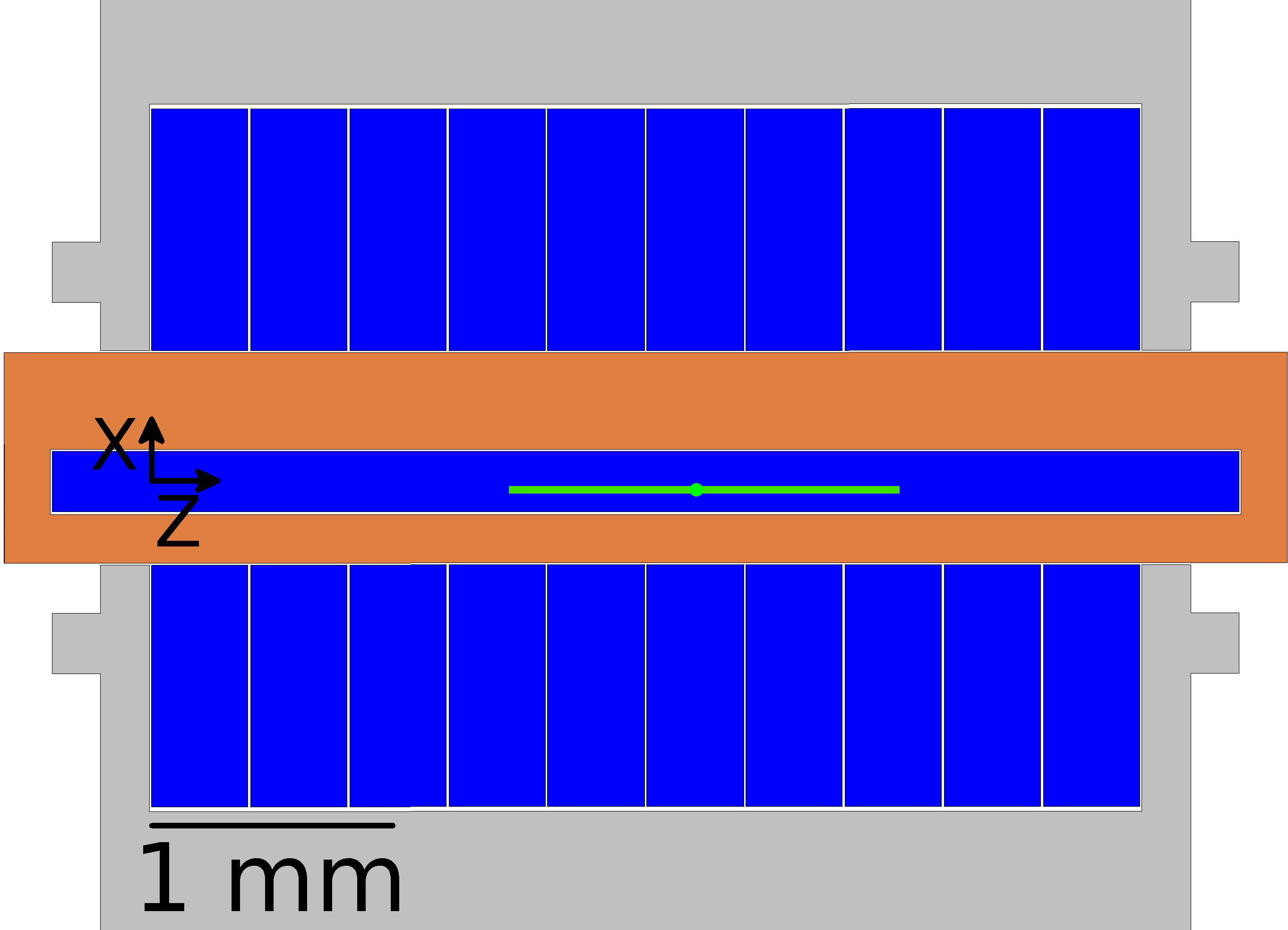}}
\subfigure{(b)\;\includegraphics[width=0.3\textwidth]{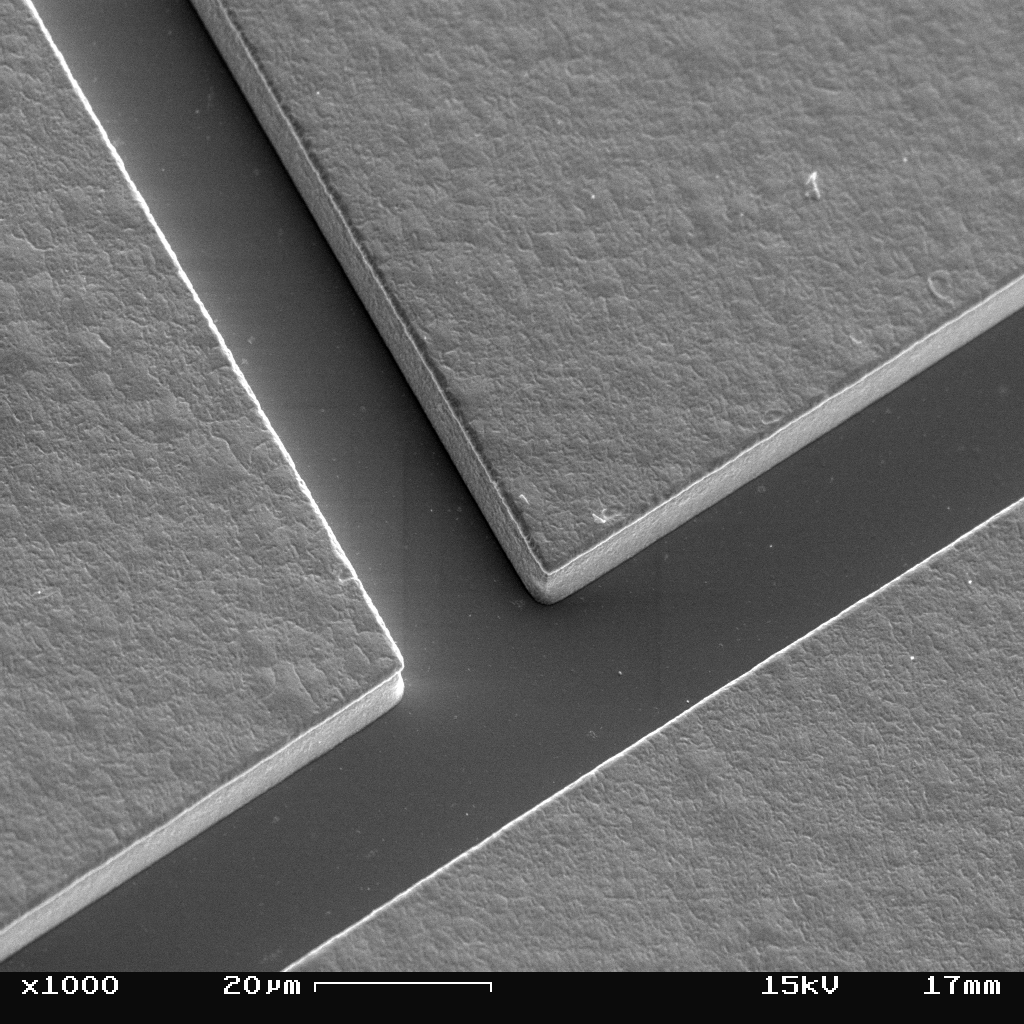}}
\end{center}
\caption{\label{fig:trap}
 (a) Simplified schematic of the trap on which extensive measurements were made. Details of the electrodes used for wire bonding to the electrodes on the periphery of the chip are not shown for simplicity. The ground electrode is shown in grey, radio-frequency electrode in orange, and DC electrodes in blue. The gaps between electrodes in this trap are 10\,$\mu$m wide. The center of the coordinate system is indicated by the axes. The green line along the $z$ axis on the central DC electrode indicates the range of axial positions in which the heating rates shown in Fig.~\ref{fig:alongtheaxislog} were measured. The circular mark on this line indicates the location used as a loading region, around which the highest increase in heating rates was observed. (b) Detail of the trap electrodes for a trap similar to the one which heating rates were made. The electrode thickness is approximately 5\,$\mu$m, and the gap between electrodes in this case is 20\,$\mu$m. The darker shade of the top surface of the trap is due to the presence of a thin layer of silver on the trap, deposited on this trap to reduce  charging of the dielectric substrate in the SEM.}
\end{figure}

We trap \Ca ions in an asymmetric surface electrode trap with trapping height of approximately 240~$\mu$m, a schematic of which is shown in Fig.~\ref{fig:trap}(a). A resistive oven produces a beam of neutral calcium atoms propagating parallel to the trap plane and along the axis of the trap. Great care has been taken to minimize direct exposure of trap surfaces to the atomic beam: A stainless steel block, elevated by 200~$\mu$m above the plane of the trap, shields the trap from the calcium produced by the oven. Measurements of fluorescence of neutral calcium produced by the oven beam reveal undetectably low fluorescence level at a height of 100~$\mu$m above the trap plane. The Ca atoms are photo-ionized in the trapping region with a two-photon process from the ground state to the continuum via the 4$^1$P$^0_1$ state using 250~mW/cm$^2$ of laser light at 422~nm (first stage) and 750~mW/cm$^2$ at 375~nm (second stage). Both laser beams are focused to waists of approximately 100~$\mu$m.

The trap is driven at frequency $\Omega_{\rm RF}/2\pi\approx$15~MHz. This drive is amplified by a radio frequency amplifier to $\approx 100\,$mW and stepped up via a helical resonator in a quarter-wave configuration to a voltage of amplitude typically around 100~V. This is fed to the RF electrodes of the trap,  with capacitance at the RF feedthrough of approximately 10~pF. DC electrodes placed on either side of the RF  electrodes are used to move the ion along the axial direction and to compensate micromotion \cite{Narayanan2010}.  DC voltages used for trapping are between -10~V and 15~V, resulting in trap frequencies of  $(f_{\rm x},\,f_{\rm y},\,f_{\rm z}) \approx (1.2,\,1.4,\,0.4)$~MHz in the horizontal radial, vertical radial,  and axial directions respectively. When the Doppler cooling lasers are switched off, ions typically remain in the trap between 5~s and 30~s depending on whether a high or low heating rate region of the trap is used.

A diode laser stabilized at 794~nm is frequency doubled using a ring cavity to produce a wavelength of 397~nm, used for Doppler cooling and detection of the ions. A second diode laser at 866~nm acts as a repump. Both lasers are frequency stabilized to within 100~kHz with independent cavities using the Pound-Drever-Hall method. The frequency of the lasers can be varied by changing the cavity lengths with piezoelectric elements  and the 397~nm laser frequency can also be adjusted without affecting the cavity using an acousto optic modulator  in a double-pass configuration. The detection laser at 397~nm is intensity stabilized at the trap to a value of 38\,${\rm mW / cm^2}$. The intensity of the 866\,nm repump laser is adjusted at approximately 115\,${\rm mW / cm^2}$.

Ion fluorescence is collected by a microscope objective with NA=0.27. It is simultaneously detected by a photomultiplier tube  and a CCD camera, using a 9:1 beam splitter. This configuration yields on the photomulitplier a maximum of 110~kcounts/s, which  is reduced to 50~kcounts/sec after the cooling laser powers have been adjusted. 

\section{Planar trap design}
\label{sec:trapdesign}

For our heating rate measurements, we use a planar trap with asymmetric radio frequency electrode. The ratio between the widths of the two RF rails is 2:1 (see Fig.~\ref{fig:trap}). This geometry gives rise to a tilted radio frequency quadrupole with tilt angle with respect to the vertical ($Y$) direction of approximately 25$^\circ$, which allows efficient laser cooling of  both radial modes. In this configuration standard Paul trap operation conditions can be achieved by aligning the static potential quadrupole with the radio frequency quadrupole. Then ion motion along each of the trap axes is described by the usual Mathieu equations \cite{Leibfried2003}. If the static and radio frequency quadrupoles are not tilted by the same amount, the result is coupling of the Mathieu equations for ion motion in different directions which could modify trap stability \cite{Shaikh2010}. This complication will arise in planar traps with symmetric radio frequency rails, since in this case tilting the trap axes with respect to the vertical direction will always result in a tilt of the static quadrupole with respect to the radio frequency quadrupole and consequently coupled equations of motion for the ions.

The trap used in this work was designed with the goal of having several independent trapping regions on the same device. Thus it has 10 pairs of DC electrode segments on the two outer sides of the radio frequency electrodes (see Fig.~\ref{fig:trap}). This gives enough freedom to trap in different regions of the trap and to transport ions along the trap axis. A central DC segment is located directly underneath the ion, facilitating micromotion compensation in the vertical direction \cite{Narayanan2010}. The central segment is 250~$\mu$m wide, while the DC segments on the sides of the radio frequency electrodes are 1000~$\mu$m by 400~$\mu$m in the $X$ and $Z$ directions respectively. The widths of the radio frequency electrodes are 400~$\mu$m and 200~$\mu$m respectively.

\section{Trap fabrication and preparation}

We fabricated this design with a minimum dimension (gaps between trap electrodes) of 10\,$\mu$m, and electrode thickness of approximately 5\,$\mu$m. The trap electrodes are deposited on a sapphire substrate which has rms surface roughness less than 1\,nm. The initial step is evaporation of a 5\,nm thick titanium adhesion layer, followed by evaporation of a 100\,nm thick gold layer. Evaporation of both is done uniformly on the substrate. Photolithography is carried out using a negative photoresist (micro resist technology, ma-N 440), suitable for deposition of resist with thickness of several micrometers. After the development step (micro resist technology ma-D 332S), the photoresist is removed from the parts of the substrate where trap electrodes will be deposited and only the regions where the sapphire substrate will be exposed are covered by photoresist of approximately 4.5\,$\mu$m thickness.

The evaporated gold acts as a seed layer during electroplating using a commercial gold electroplating solution (Metalor, ECF 60) at a temperature of 50$^o$\,C and current density of 1.4\,mA/cm$^2$. The resulting deposition rate is approximately 75\,nm/min and the process is carried on for 70\,minutes. The resulting gold layer has thickness varying between 4.4\,$\mu$m and 6.3\,$\mu$m on different areas of a circular 50\,mm diameter wafer. The surface rms roughness of the gold film as measured with a surface profilometer is approximately 17\,nm. After electroplating, the photoresist is removed and the seed layer is etched away using diluted ``aqua regia'', with HCl:HNO$_3$:H$_2$O in a 3:1:7 ratio. In this process etching of the seed layer in the regions of the gaps between trap electrodes will proceed at a rate considerably slower than in the rest of the trap, due to insufficient wetting of high aspect ratio troughs by the etching solution. To improve the etching rate of the seed layer, we use a surfactant (Triton X-100). Finally, after removal of the gold seed layer, a longer titanium-specific etching step to remove the remains of the titanium adhesion layer is performed in a diluted hydrochloric acid solution (32\% HCl with addition of Triton-X 100, heated to 50$^o$\,C). The final etching steps do not influence the surface roughness of the trap electrodes. Before installation in vacuum all traps are cleaned in a sequence of acetone/isopropyl alcohol/deionized water baths, and blow dried in a nitrogen gas stream.

Using the above procedure, we have fabricated traps of various sizes. The ion trapping height in these varies between 500\,$\mu$m and 125\,$\mu$m. We successfully trapped single ions in three of those traps, the first with ion height of 500\,$\mu$m above the trap surface, and the other two with ion height of approximately 240\,$\mu$m. For what follows, we focus on one of the latter, which was extensively studied. Prior to the heating rate measurements, this trap was baked in vacuum at 150$^\circ$C for two weeks, to reach a base pressure of $2\times10^{-10}\,$mbar.

\section{Heating rate measurement method}
\label{sec:heat-rates}

The heating rates of the trap have been measured using the Doppler recooling method in which the motional energy of the ion is determined by taking advantage of the increased Doppler shift for an ion with increased kinetic energy \cite{Wesenberg2007}. This allows one to estimate the ion energy by comparing the instantaneous ion fluorescence to the steady state fluorescence level of a Doppler cooled ion. To evaluate the change in fluorescence, Doppler cooling of the ion is turned off for time $\tau_{\rm off}$ during which time the ions heat up. The change in the ion fluorescence is monitored as the ion cools down after turning the Doppler cooling on. This change in fluorescence level is used to determine the energy that the ion acquired during $\tau_{\rm off}$, and thus deduce the ion heating rate.

Application of the Doppler recooling method as described in the original proposal is practical for situations where the ion mode which is Doppler cooled does not experience significant micromotion, for example to cooling of the axial mode of a linear Paul trap. While this geometry is frequently sufficient, it is not applicable whenever a single Doppler cooling beam is available, as in situations where planar traps are part of a complex setup granting limited optical access to the ions. In these cases micromotion contribution to Doppler broadening can be significant and has to be taken into account in determining ion energy from changes in fluorescence. Following the procedure discussed in \cite{Wesenberg2007} we numerically find that for our experimental parameters micromotion contributes significantly to Doppler broadening, and can lead to an overestimation of the ion energy by a factor between 2 and 4 if the method is applied without accounting for micromotion. Nevertheless, the numerical integration required to obtain and fit fluorescence recooling curves to experimental data is computationally intense, which makes it practically infeasible. 

In the case of \Ca ions, further complications to application of the Doppler recooling method arise due to the electronic level scheme  which is more comlicated than that of a simple two-level atom. In order to estimate the ion energy, the linewidth of the resonance and the saturation parameter(s) are required. For two level systems this is straightforward, however for multi-level dynamics the situation  is more complex. In particular, dark-resonances and the strength of the repump light field(s) have significant influence on the line shape \cite{Lindberg1986,Siemers1992,Reiss1996}.

One possible resolution is the detailed modeling of the dynamics and careful measurement of all relevant laser parameters such as detunings and intensities. Nevertheless, due to the complexity of the situation, instead of modelling the recooling dynamics in the presence of micromotion and multiple atomic levels we use here a simple workaround by calibrating the energy scale with well-defined electric field noise. Apart from ease of implementation a further advantage is that one can separate out the effect of different energies of the three motional modes on the fluorescence signal by choosing particular noise frequency bands and/or excitation geometries, i.e. the electrodes to which the noise is applied. Thus, one can study or avoid implicit assumptions while extracting a single heating rate for the three motional modes.

\begin{figure}[heating]
\subfigure{\includegraphics[width=0.7\textwidth]{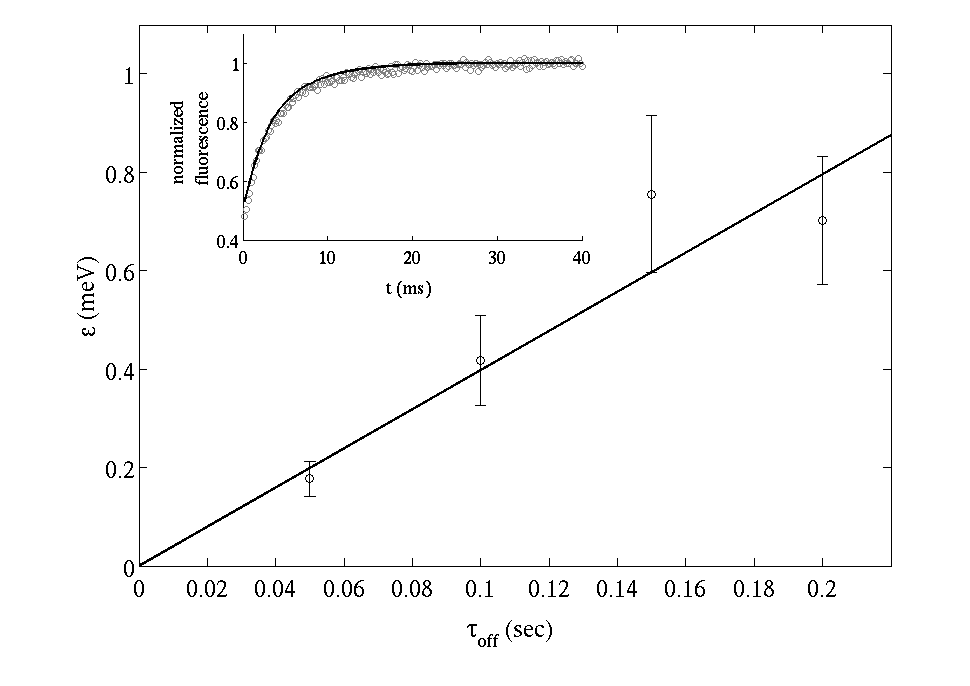}}
\caption{\label{fig:pvso}Scaled energy $\epsilon$ as a function of ion heating time $\tau_{\rm off}$ in Doppler recooling experiments. As expected, the energy of the ion scales linearly with heating time and a linear fit constrained through the origin is used to derive the rate of change $d\epsilon/dt$, in this  case $4.0\pm0.3$~meV/s. Uncertainties of the ion energy are determined by uncertainty in the exact value of the detuning used for the measurement \textit{Inset}:  A typical Doppler recooling curve for one value of $\tau_{\rm off}$ with fluorescence as a function of time, and fit (solid line) according to the procedure described in \cite{Wesenberg2007}. Time zero in the inset corresponds to onset of recooling, i.e. turning on of the repump laser at 866\,nm. }
\end{figure}

For these experiments, the intensity of the detection laser at 397\,nm is actively stabilized at a value of 38\,${\rm mW / cm^2}$. The frequency of the detection laser is detuned to 5\,MHz on the red side of the resonance of $S_{\frac{1}{2}}\leftrightarrow P_{\frac{1}{2}}$. The intensity of the 866\,nm repump laser is adjusted at 115\,${\rm mW / cm^2}$, and its frequency is adjusted to the  $P_{\frac{1}{2}}\leftrightarrow D_{\frac{3}{2}}$ transition by maximizing the fluorescence level.  To measure recooling curves, we turn off Doppler cooling and allow the ion to heat by switching off the repump laser at 866\,nm for the variable time $\tau_{\rm off}$. During a time interval $t_{\rm pump}\sim 0.1\, \mu{\rm s} \ll \tau_{\rm off}$  the ion gets pumped into the $D_{\frac{3}{2}}$ level and will not be cooled by the Doppler cooling laser. After the repump laser is turned on, ion fluorescence is acquired using time bins of $50~\mu s$. The procedure is repeated typically 1000 times and the fluorescence curves are averaged to improve statistics. The inset in Fig.\,\ref{fig:pvso} shows a typical recooling curve obtained using this procedure.

To extract the ion energy from the recooling curves, we use the analysis procedure of \cite{Wesenberg2007} and extract a quantity $\epsilon$, to which we will refer as the scaled energy. This would correspond to the ion energy in the case of a two-level ion experiencing no micromotion, but in our case it is found to be merely proportional to the ion energy. The latter is evident in the linear increase of $\epsilon$ with heating time $\tau_{\rm off}$, shown in Fig.~\ref{fig:pvso}.

To calibrate the scaled energy, we heat the ion with externally supplied white electric field noise provided  by a signal generator that has been calibrated against a spectrum analyzer. The noise is white to a very large degree,  with spectral density at the ion secular frequencies varying by less than  10~$\%$. We vary the noise spectral density between $4.4\times10^{-16}\,\rm{ V}^2/\rm{Hz}$ and $2.5\times10^{-14}\,\rm{V^2}/\rm{Hz}$. The noise is applied  to one of the trap electrodes with almost equal projection of the electric field on all the three modes of ion motion and the corresponding electric field noise at the ion position is changed between $1.7\times10^{-11}\,\rm{(V/m)}^2/\rm{Hz}$  and $1.0\times10^{-9}\,\rm{(V/m)}^2/\rm{Hz}$. 

\begin{figure}[heating]
\subfigure{\includegraphics[width=0.7\textwidth]{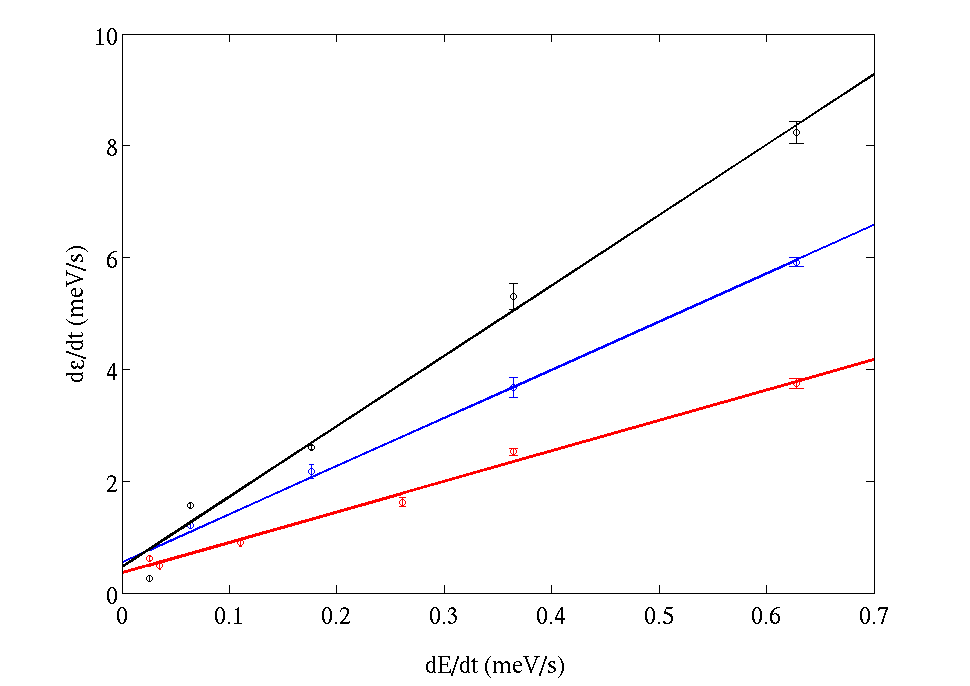}}
\caption{\label{fig:mesEvsindE}Calibration of the scaled energy, $\epsilon$, determined from recooling curves against the externally applied noise source. Shown is $d\epsilon/dt$, determined from the procedure outlined in the text and in Fig.\,\ref{fig:pvso} versus the externally induced heating rate $dE/dt$, determined from Eq.~\ref{eqn:ndot}, for different magnitudes of applied noise. Also shown are a linear fits to the data. Different curves were measured with a minimum separation of 4 hours.}
\end{figure}

The calibration is performed by determining the heating rate $d\epsilon/dt$ from the recooling curves and  comparing it to the expected ion heating rate $dE/dt$, at various levels of applied external noise. The latter is calculated  using \cite{Turchette2000}
\begin{eqnarray}
\label{eqn:ndot}
E=\sum_{i=\{x,y,z\}} \frac{e^2}{4m}\,S_E(\omega_i)\,\tau_{\rm off}\;,
\end{eqnarray}
where $\tau_{\rm off}$ is the time the ion is allowed to heat before measuring its energy, $\omega_i$ is the secular frequency of the $i^{\rm th}$ mode of the ion $S_E(\omega_i)$ is the power spectral density of the electric field at that frequency. In this expression heating of the secular sidebands of the micromotion at $\Omega_{RF}\pm\omega_i$ is omitted, since it is expected to be weaker by a factor $(\omega_i/\Omega_{\rm RF})^2\sim0.01$ which is significantly below the accuracy of this method. Heating rates are determined  for different values of the externally applied noise. The results of such measurements are shown in Fig.~\ref{fig:mesEvsindE}, where we plot $d\epsilon/dt$ versus $dE/dt$. We find linear dependence on time scales about one hour, but the slopes  between different trends taken several hours apart vary by up to a factor of roughly 2. This is most likely due to drifts in the experimental parameters  that we do not stabilize accurately, namely the repump laser detuning and intensity. From linear fits to these curves we determine $\epsilon$  to be a factor of $8.9\pm3.6$ higher than the actual energy acquired by the ion. As a result, the uncertainty in our determination of heating rates of our trap is ultimately determined by the uncertainty in the calibration of $\epsilon$ vs $E$.

\section{Results and discussion}

\begin{figure}[heating]
\subfigure{\includegraphics[width=0.7\textwidth]{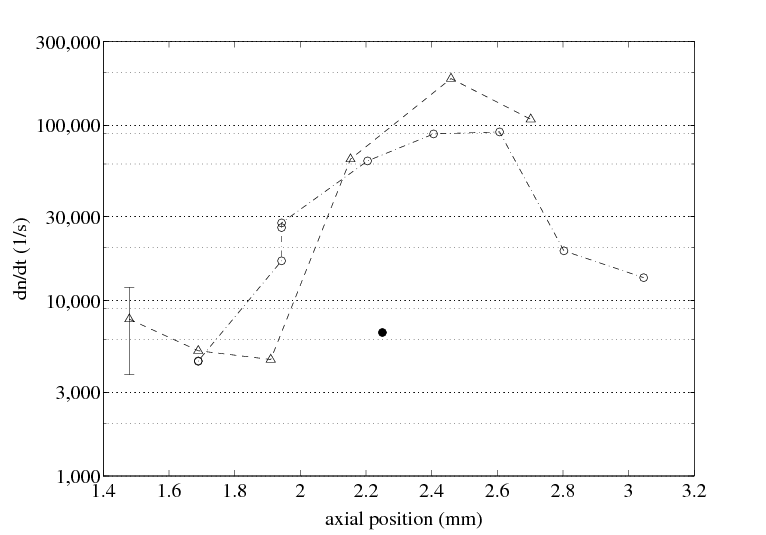}}
\caption{\label{fig:alongtheaxislog} Heating rate, normalized to 1~MHz phonons, at different positions along the axis of the trap and at different times. A representative errorbar is shown on the data point at position 1.5~mm. The filled data point at 2.25~mm was measured in the ``pristine'' trap, i.e. after 1.5 months of operation in vacuum.  We mark the measurements performed 8 and 10 months after the initial installation of the trap in vacuum with the open black triangles and open black circles respectively. At the time of the latter measurements the heating rate around the ion loading region had increased by one order of magnitude, as discussed in the text.}
\end{figure}

Heating rate measurements were performed at different positions along the axis of the trap, as shown in Fig.~\ref{fig:alongtheaxislog}. The measured results have been normalized to phonons of 1~MHz frequency. The single filled data point at axial position  of 2250~$\mu{\rm m}$ was measured one and a half months after the installation of the trap  in vacuum and the open data points were obtained later, after an abrupt increase in heating rates of our trap. This occurred approximately after two months of trap operation. It coincided with a failure of the ion pump used to maintain the ultra-high vacuum and was followed by an increase in the stray electrostatic fields at the ion positions \cite{Narayanan2010}. During the pump failure elevated pressures of $10^{-7}$~mbar were reached in the vacuum chamber, while before and after the pressure was in the low $10^{-10}~{\rm mbar}$ range.

The heating rate appears to have a maximum around the region used for ion loading in the trap. This is the range of axial positions between roughly 1900~$\mu$m and 2700~$\mu$m, which was used extensively as an ion-loading zone. Far from the loading region the heating rate is close to the values measured in the ``pristine'' trap $1\,\frac{1}{2}$ months after trap installation in vacuum. The heating rates near the loading zone of the trap increased from $\approx 5$~phonons/ms to more than $\approx 50$~phonons/ms in the first few months of trap operation. Far from this zone, the heating rates maintained their lower value. Likewise, uncooled  ion lifetimes decreased by one order of magnitude in the parts of the trap where increased heating  rates were observed.

The change in heating rates coincided with an abrupt increase in stray static fields in the same region of the trap \cite{Narayanan2010} and it is very likely that both of these effects are related to extensive use of the region around axial coordinate of 2250~$\mu$m as a loading zone. A number of effects can be responsible for the change in trap behavior. From the geometry of our vacuum apparatus, we expect that possible Ti contamination from the titanium sublimation pump is rather homogeneous. Similarly, surface contamination with atomic Ca from the oven is not expected to peak at the loading zone, located near the center of the trap, but rather to fall off monotonically moving towards the oven direction as a result of the atomic beam flux screening, discussed in Sec.~\ref{sec:setup}. One possible meachanism of surface contamination is bombardment of the trap electrodes by electrons created during trap loading or by \Ca ions which are created by photoionization outside the trapping volume. These can impinge on the gold surface with energies up to 100~eV when accelerated by the RF field. \Ca ions with energies of a few eV can get physisorbed on the trap electrodes and later form chemical  compounds, while ions at higher energies can sputter material from the trap surface. In addition, it is  possible that the laser light used for ion creation and detection locally alters the chemical composition.

As the mechanism leading to heating of trapped ions is not fully understood, it is useful to summarize heating rates in different traps constructed from different electrode materials and used to trap different ion species. Since the noise spectral density, $S_E(\omega)$, causing anomalous heating is believed to scale as $1/f$ with frequency\cite{Turchette2000}, it is convenient to regularize experimental results by deriving $\omega_{i}\,S_E(\omega_{i})$ \cite{Epstein2007} where $\omega_{i}$ is the frequency of the ion mode which is being measured. This is straightforwardly done for heating rate measurements performed using the sideband spectroscopy method, where the heating rate of a specific secular mode can be measured. In order to regularize results measured using the Doppler recooling method one needs to take into account that the energy of all three secular modes is redistributed and finally removed from the ion, as discussed in \cite{Wesenberg2007}. Thus, a properly averaged ``effective'' frequency has to be used. By energy conservation considerations, we find that the effective frequency is $\bar{\omega}^{-1}=\sum_{i=1}^3{\omega_{i}^{-1}}$. In our measurements, the axial frequency is significantly smaller than the radial ones and $\bar{\omega}$ turns out to be very close to the axial frequency.

\begin{figure}[heating-summary]
\includegraphics[width=\textwidth]{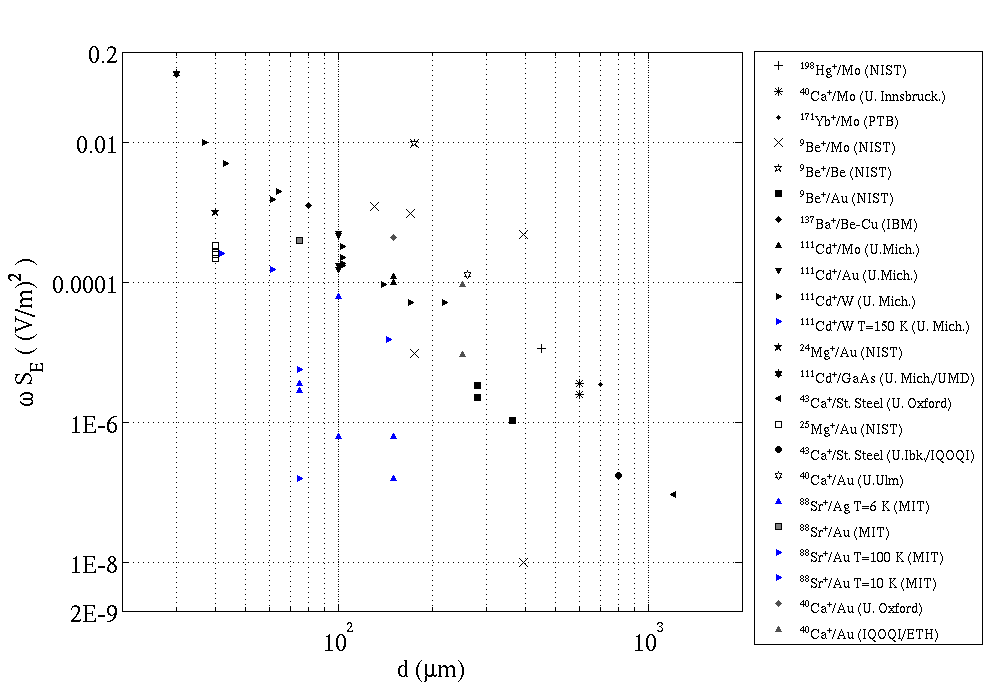}
\caption{\label{fig:heat-rate-summary} Summary of representative noise spectral densities for different traps found in the literature, plotted versus ion-trap distance $d$. The ion species and trap electrode material are specified. Temperature is specified when different than room temperature. References are organized in chronological order of appearance. The trap used in this work fell well within the average trend initially (lower point, corresponding to heating rate measured $1.5$ months after installing the trap in vacuum), but still falls within the limits of the trend even after deterioration of trap performance. The diagonal in the figure corresponds to a $d^{-4}$  trend. Data are from references  \cite{Seidelin2006,Labaziewicz2008,Allcock2009,Monroe1995,Turchette2000,DeVoe2002,Labaziewicz2008a,Epstein2007,Diedrich1989,Roos1999,Tamm2000,DesLauriers2004,DesLauriers2006a,Stick2006,Lucas2007,Schulz2008a}.}
\end{figure}

Using this approach, we have compiled a graph of representative data found in the literature \cite{Seidelin2006,Labaziewicz2008,Allcock2009,Monroe1995,Turchette2000,DeVoe2002,Labaziewicz2008a,Epstein2007,Diedrich1989,Roos1999,Tamm2000,DesLauriers2004,DesLauriers2006a,Stick2006,Lucas2007,Schulz2008a}, see Fig.~\ref{fig:heat-rate-summary}. In the graph we specify ion species used in the measurements and trap electrode material, as well as temperature of the trap, if other than room temperature. We include some of the measurements performed in cooled trap apparatuses, but focus on the room temperature results where measurements are abundant. We observe that the majority of traps measured at the same temperature fall around a general trend following a $d^{-4}$ law (corresponding to the diagonal of the figure). The traps with the lowest heating rates have been a molybdenum ring trap \cite{Turchette2000} used to trap $^{9}{\rm Be}^{+}$ and a gold-on-quartz surface trap \cite{Epstein2007} used to trap $^{25}{\rm Mg}^{+}$. Nevertheless, the presence of both gold and molybdenum traps with noise spectral densities well within the average implies that more than material-specific the source of the noise is specific to trap preparation and cleaning procedures. As we discuss in what follows, less than one monolayer of adsorbed impurity atoms, as well as a thin dielectric layer formed on the trap surface can be responsible for the observed heating rates.

\section{Model for anomalous heating}
\label{sec:heat-model}

We now proceed to derive a phenomenological model of ion heating that can provide physical information to help identify the possible sources of anomalous heating in ion traps. The model is required to reproduce the correct scaling of noise spectral density with ion-trap distance and with frequency. Thus we consider a distribution of fluctuating dipoles on the trap electrode surface to be the source of anomalous heating. In many situations $1/f$ noise is attributed to systems with a Debye-type relaxation, and a distribution of relaxation rates \cite{Dutta1981}. Accordingly, we consider a typical dipole with dipole moment $\mu$, relaxation rate $\Gamma$, leading to an autocorrelation function
\begin{equation}
\phi_{\mu}(t) = \mu^2\,e^{-\Gamma\,t}
\end{equation}
The corresponding dipole spectral density is
\begin{equation}
S_{\mu}^{\Gamma}(f) = \frac{2\,\mu^2\,\Gamma}{\Gamma^2+(2\pi f)^2}
\end{equation}
To recover the $1/f$ noise spectrum, we assume a distribution of relaxation rates scaling as $F(\Gamma) = A/\Gamma$,  where the normalization $A = \ln\left(\frac{\Gamma_{\rm max}}{\Gamma_{\rm min}}\right)$ depends only logarithmically on the limiting relaxation rates. In the context of thermally activated dipole relaxation $A$ corresponds to the range of energy barriers through which relaxation can occurr. The dipole moment spectral density is now obtained
\begin{equation}
S_{\mu}(f) = \frac{A\,\mu^2}{2\,f}
\end{equation}

Fluctuating dipoles on the trap electrodes produce fluctuating electric fields of order $E = \frac{\mu}{2\,\pi\,\epsilon_0\,r^3}$ at the ion position, a distance $r$ away. This relation will not be exact, since a fluctuating dipole near a conducting surface will get either enhanced or screened depending on its orientation \cite{Sommerfeld1949}, and the exact mechanism of screening of charges by metal surfaces at separations of the order of  interatomic distances is unclear \cite{Feibelman1982}, but it will suffice for our current estimates. Moreover, fluctuating electric dipoles will produce negligible magnetic field noise. The reason is that in the near-field of dipole radiation the electromagnetic field is predominantly electric.  The near-field condition is always satisfied in current experiments, so this mechanism will cause electric field noise and virtually no magnetic field noise.

To estimate the electric field noise spectral density, we perform statistical averaging which results in scaling as $\sqrt{N}$ with the number, $N$, of dipoles. The noise spectral density from a planar trap with surface density $n_{\rm S}$ of dipoles is then given by the integral over the trap surface 
\begin{equation}\label{eq:spectral-density-final}
S_{E}(f) = n_{s}\,\int_{\rm surf}\left(\frac{1}{2\,\pi\,\epsilon_0\,r}\right)^2\,S_{\mu}(f)\,d\alpha =
\frac{A\, n_{\rm S}\, \mu^2}{8\, \pi\, \epsilon_{0}^2\, d^4\, f}
\end{equation}
The rudimentary result of Eq.~\ref{eq:spectral-density-final} allows us to estimate the product $n_{\rm S}\,\mu^2$ of the density of dipoles on the trap surface with the squared dipole moment, based on the measured noise spectral density, secular frequency, and ion-trap distance. We use this result to evaluate different possible sources of anomalous heating due to contaminated trap electrode surfaces. By assuming a typical dipole moment of 1~D and value of normalization constant $A\approx10$ as reasonable estimates, the value of the average trend of heating rate data in Fig.~\ref{fig:heat-rate-summary} leads to $n_{\rm S}\approx 6\times10^{19}\,/{\rm m}^2$. This value corresponds to typical surface densities of atoms on solid surfaces.

First we consider impurity atoms adsorbed on the trap electrodes, following the analysis of Volokitin \textit{et. al.} \cite{Volokitin2005}. In this case, a metal surface with small coverage by adsorbed molecules can give rise to dissipation and noise several orders of magnitude higher than those expected from a clean metal surface. To put this in the context of our model, we consider adsorbed atoms on the metal surface which form polar bonds and can switch between different configurations giving rise to fluctuating dipoles. As a typical dipole moment we take the value for Cs adsorbed on Cu(100) as considered by the above authors in \cite{Volokitin2005} i.e. $\mu\approx$~4~D. Using Eq.~\ref{eq:spectral-density-final}, we obtain a coverage fraction $\theta\approx0.6$ for a disordered gold surface, which is a very reasonable result. While in the case of tunneling TLS, which we consider next, the typical ranges of activation energies and relaxation times are somewhat known and well understood, such details are in general unclear for surface adsorbates. Nevertheless, this mechanism  requires less than one monolayer of adsorbate and can, as a result, be very common unless specific care is taken to target  and remove material-specific adsorbates from metal surfaces.

We also consider tunneling TLS in an amorphous dielectric film on the trap electrodes as the fluctuating dipoles. The thickness of the dielectric layer can be estimated as $\delta = n_{\rm S}/n_{\rm V}$, where $n_{\rm V}$ is the density of TLS in the dielectric. While this quantity is relatively well determined at temperatures in the few K range \cite{Phillips1987}, the situation at higher temperatures is unclear. It has been argued that the TLS density of states increases linearly with temperature \cite{Varma1982}, and according to that estimate one obtains at room temperature a density $n_{\rm V}\approx 5\times10^{27}\,/{\rm m}^3$ of TLS with energies less than the estimated upper limit of $0.1~\rm{eV}$ \cite{Anderson1972,Phillips1972}. With the above value of $n_{\rm s}$, this would correspond to a dielectric of thickness $\delta\approx10\,{\rm nm}$. While this is a high value, it is only intended as an order of magnitude estimate, given the crudeness of the approximations used. It is possible that oxides on metal surfaces, residues of fabrication, gases adsorbed on the trap electrodes, and material deposited on trap surfaces while the trap is in operation form dielectric layers in the nanometer range. 

\section{Summary and Conclusions}

We reported a simple fabrication method for segmented planar ion traps, suitable for ion transport and for operation with a tilt of the main trap axes with respect to the direction perpendicular to the trap plane. We described a modification of the Doppler recooling method for heating rate measurements and used the method to measure heating rates of one gold-on-sapphire planar trap. The heating rates were found to deteriorate with time, and to be position dependent with a maximum around the loading zone of the trap, suggesting surface contamination as a major source of anomalous heating. We presented a phenomenological model that can explain anomalous heating in the ion traps and used it to evaluate different possible sources of anomalous heating encountered in other systems. We found that both adsorbed impurity atoms close to a density of one monolayer and tunneling TLS in amorphous dielectrics on the trap electrodes  can be responsible for the observed heating. Careful surface characterization and cleaning procedures of ion traps are needed to have a more definite  conclusion on the sources of and solutions to anomalous heating.

\section*{Acknowledgments}

The authors would like to thank John Martinis for useful discussions and D. Frank Ogletree for pointing out important work on non-contact friction. This work was supported by the Austrian Ministry of Sciences with a START grant. N.~Daniilidis was supported in part by the European Union with a Marie Curie fellowship. S.~Narayanan was partially supported by the Laboratory Directed Research and Development Program of Lawrence Berkeley National Laboratory under U.S. Department of Energy Contract No.~DE-AC02-05CH11231, and St.~Schulz was supported by the Alexander von Humboldt Foundation with a Feodor Lynen Fellowship.

\newpage

\bibliographystyle{unsrt}
\bibliographystyle{abbrv.bst}

\begin{thebibliography}{10}

\bibitem{Haeffner2008}
H~H\"{a}ffner, C~F Roos, and R~Blatt.
\newblock {Quantum computing with trap ions}.
\newblock {\em Physics Reports}, 469:155, 2008.

\bibitem{Blatt2008}
R~Blatt and D~Wineland.
\newblock {Entangled states of trapped atomic ions}.
\newblock {\em Nature}, 453:1008--1015, 2008.

\bibitem{Kielpinski2002}
D~Kielpinski, C~Monroe, and D~J Wineland.
\newblock {Architecture for a large-scale ion-trap quantum computer.}
\newblock {\em Nature}, 417(6890):709--711, 2002.

\bibitem{Chiaverini2005}
J.~Chiaverini, R.~B. Blakestad, J.~Britton, J.~D. Jost, C.~Langer,
  D.~Leibfried, R.~Ozeri, and D.~J. Wineland.
\newblock {Surface-electrode architecture for ion-trap quantum information
  processing}.
\newblock {\em Quantum Information and Computation}, 5:419--439, 2005.

\bibitem{Seidelin2006}
S~Seidelin, J~Chiaverini, R~Reichle, J~J Bollinger, D~Leibfried, J~Britton, J~H
  Wesenberg, R~B Blakestad, R~J Epstein, D~B Hume, W~M Itano, J~D Jost,
  C~Langer, R~Ozeri, N~Shiga, and D~J Wineland.
\newblock {Microfabricated surface-electrode ion trap for scalable quantum
  information processing.}
\newblock {\em Physical Review Letters}, 96:253003, 2006.

\bibitem{Britton2006}
J~Britton, D~Leibfried, J~Beall, R~B Blakestad, J~J Bollinger, J~Chiaverini,
  R~J Epstein, J~D Jost, D~Kielpinski, C~Langer, R~Ozeri, R~Reichle,
  S~Seidelin, N~Shiga, J~H Wesenberg, and D~J Wineland.
\newblock {A microfabricated surface-electrode ion trap in silicon}.
\newblock {\em arXiv:quant-ph/0605170}, 2006.

\bibitem{Pearson2006}
C~E Pearson, D~R Leibrandt, W~S Bakr, W~J Mallard, K~R Brown, and I~L Chuang.
\newblock {Experimental investigation of planar ion traps}.
\newblock {\em Physical ReviewA}, 73:32307, 2006.

\bibitem{Labaziewicz2008}
J~Labaziewicz, Y~Ge, P~Antohi, D~Leibrandt, K~R Brown, and I~L Chuang.
\newblock {Suppression of Heating Rates in Cryogenic Surface-Electrode Ion
  Traps}.
\newblock {\em Physical Review Letters}, 100:13001, 2008.

\bibitem{Leibrandt2009}
D~R Leibrandt, J~Labaziewicz, R~J Clark, I~L Chuang, R~Epstein, C~Ospelkaus,
  J~Wesenberg, J~Bollinger, D~Leibfried, D~Wineland, D~Stick, J~Sterk,
  C~Monroe, C.-S. Pai, Y~Low, R~Frahm, and R~E Slusher.
\newblock {Demonstration of a scalable, multiplexed ion trap for quantum
  information processing}.
\newblock {\em arXiv:0904.2599v1 [quant-ph]}, 2009.

\bibitem{Allcock2009}
D.~T.~C. Allcock, J.~A. Sherman, M.~J. Curtis, G.~Imreh, A.~H. Burrell, D.~J.
  Szwer, D.~N. Stacey, A.~M. Steane, and D.~M. Lucas.
\newblock {Implementation of a symmetric surface electrode ion trap with field
  compensation using a modulated Raman effect}.
\newblock {\em arXiv:0909.3272v2}, 2009.

\bibitem{Amini2009}
J.~M. Amini, H.~Uys, J.~H. Wesenberg, S.~Seidelin, J.~Britton, J.~J. Bollinger,
  D.~Leibfried, C.~Ospelkaus, A.~P. VanDevender, and D.~J. Wineland.
\newblock {Scalable ion traps for quantum information processing}.
\newblock {\em arXiv:0812.3907v1}, 2009.

\bibitem{Tian2004}
L~Tian, P~Rabl, R~Blatt, and P~Zoller.
\newblock {Interfacing quantum-optical and solid-state qubits.}
\newblock {\em Physical Review Letters}, 92(24):247902, June 2004.

\bibitem{Daniilidis2009}
N~Daniilidis, T~Lee, R~Clark, S~Narayanan, and H~H\"{a}ffner.
\newblock {Wiring up trapped ions to study aspects of quantum information}.
\newblock {\em Journal of Physics B}, 42:154012, 2009.

\bibitem{Monroe1995}
C~Monroe, D~M Meekhof, B~E King, W~M Itano, and D~J Wineland.
\newblock {Demonstration of a fundamental quantum logic gate.}
\newblock {\em Physical Review Letters}, 75:4714--4717, 1995.

\bibitem{Wineland1998}
D~J Wineland, C~Monroe, W~M Itano, D~Leibfried, B~E King, and D~M Meekhof.
\newblock {Experimental Issues in Coherent Quantum-State Manipulation of
  Trapped Atomic Ions}.
\newblock {\em Journal of Research of the National Institute for Standards and
  Technology}, 103:259--328, 1998.

\bibitem{Turchette2000}
Q~A Turchette, Kielpinski, B~E King, D~Leibfried, D~M Meekhof, C~J Myatt, M~A
  Rowe, C~A Sackett, C~S Wood, W~M Itano, C~Monroe, and D~J Wineland.
\newblock {Heating of trapped ions from the quantum ground state}.
\newblock {\em Physical Review A}, 61:63418, 2000.

\bibitem{DeVoe2002}
Ralph DeVoe and Christian Kurtsiefer.
\newblock {Experimental study of anomalous heating and trap instabilities in a
  microscopic $^{137}$Ba ion trap}.
\newblock {\em Physical Review A}, 65(6):1--8, June 2002.

\bibitem{Labaziewicz2008a}
Jaroslaw Labaziewicz, Yufei Ge, David Leibrandt, Shannon~X Wang, Ruth Shewmon,
  and Isaac~L Chuang.
\newblock {Temperature Dependence of Electric Field Noise Above Gold Surfaces}.
\newblock {\em Physical Review Letters}, 101:180602, 2008.

\bibitem{Li2007}
Mo~Li, H~X Tang, and M~L Roukes.
\newblock {Ultra-sensitive NEMS-based cantilevers for sensing, scanned probe
  and very high-frequency applications.}
\newblock {\em Nature nanotechnology}, 2(2):114--20, February 2007.

\bibitem{Mamin2003}
H~J Mamin, R~Budakian, B~W Chui, and D~Rugar.
\newblock {Detection and manipulation of statistical polarization in small spin
  ensembles.}
\newblock {\em Physical Review Letters}, 91(20):207604, November 2003.

\bibitem{Speake2003}
C~C Speake and C~Trenkel.
\newblock {Forces between Conducting Surfaces due to Spatial Variations of
  Surface Potential}.
\newblock {\em Physical Review Letters}, 90(16):160403, April 2003.

\bibitem{Zimmerli1992}
G~Zimmerli, TM~Eiles, RL~Kautz, and JM~Martinis.
\newblock {Noise in the Coulomb blockade electrometer}.
\newblock {\em Applied Physics Letters}, 61(July):237--239, 1992.

\bibitem{Astafiev2004}
O~Astafiev, Yu~A Pashkin, Y~Nakamura, T~Yamamoto, and J~S Tsai.
\newblock {Quantum noise in the josephson charge qubit.}
\newblock {\em Physical Review Letters}, 93(26):267007, December 2004.

\bibitem{Gao2008}
Jiansong Gao, Miguel Daal, Anastasios Vayonakis, Shwetank Kumar, Jonas
  Zmuidzinas, Bernard Sadoulet, Benjamin~a. Mazin, Peter~K. Day, and Henry~G.
  Leduc.
\newblock {Experimental evidence for a surface distribution of two-level
  systems in superconducting lithographed microwave resonators}.
\newblock {\em Applied Physics Letters}, 92(15):152505, 2008.

\bibitem{Kumar2008}
Shwetank Kumar, Jiansong Gao, Jonas Zmuidzinas, Benjamin~A. Mazin, Henry~G.
  LeDuc, and Peter~K. Day.
\newblock {Temperature dependence of the frequency and noise of superconducting
  coplanar waveguide resonators}.
\newblock {\em Applied Physics Letters}, 92(12):123503, 2008.

\bibitem{OConnell2008}
Aaron~D. O’Connell, M.~Ansmann, R.~C. Bialczak, M.~Hofheinz, N.~Katz, Erik
  Lucero, C.~McKenney, M.~Neeley, H.~Wang, E.~M. Weig, a.~N. Cleland, and J.~M.
  Martinis.
\newblock {Microwave dielectric loss at single photon energies and millikelvin
  temperatures}.
\newblock {\em Applied Physics Letters}, 92(11):112903, 2008.

\bibitem{Hayashi2003}
T.~Hayashi, T.~Fujisawa, H.~Cheong, Y.~Jeong, and Y.~Hirayama.
\newblock {Coherent Manipulation of Electronic States in a Double Quantum Dot}.
\newblock {\em Physical Review Letters}, 91(22):1--4, November 2003.

\bibitem{Gorman2005}
J.~Gorman, D.~Hasko, and D.~Williams.
\newblock {Charge-Qubit Operation of an Isolated Double Quantum Dot}.
\newblock {\em Physical Review Letters}, 95(9):1--4, August 2005.

\bibitem{Shnirman2005}
Alexander Shnirman, Gerd Sch\"{o}n, Ivar Martin, and Yuriy Makhlin.
\newblock {Low- and High-Frequency Noise from Coherent Two-Level Systems}.
\newblock {\em Physical Review Letters}, 94(12):1--4, April 2005.

\bibitem{Martinis2005}
John Martinis, K.~Cooper, R.~McDermott, Matthias Steffen, Markus Ansmann,
  K.~Osborn, K.~Cicak, Seongshik Oh, D.~Pappas, R.~Simmonds, and Clare Yu.
\newblock {Decoherence in Josephson Qubits from Dielectric Loss}.
\newblock {\em Physical Review Letters}, 95(21):1--4, November 2005.

\bibitem{Anderson1972}
P~W Anderson, B~I Halperin, and C~M Varma.
\newblock {Anomalous low-temperature thermal properties of glasses and spin
  glasses}.
\newblock {\em Philosophical Magazine}, 25(1):1--9, January 1972.

\bibitem{Phillips1972}
W~A Phillips.
\newblock {Tunneling States in Amorphous Solids}.
\newblock {\em Journal of Low Temperature Physics}, 7(3):351--360, 1972.

\bibitem{Phillips1987}
W~A Phillips.
\newblock {Two-level states in glasses}.
\newblock {\em Reports on Progress in Physics}, 50(12):1657, 1987.

\bibitem{Arcizet2009}
O.~Arcizet, R.~Rivi\`{e}re, a.~Schliesser, G.~Anetsberger, and T.~Kippenberg.
\newblock {Cryogenic properties of optomechanical silica microcavities}.
\newblock {\em Physical Review A}, 80(2):1--4, August 2009.

\bibitem{Dorofeyev1999}
I.~Dorofeyev, H.~Fuchs, G.~Wenning, and B.~Gotsmann.
\newblock {Brownian Motion of Microscopic Solids under the Action of
  Fluctuating Electromagnetic Fields}.
\newblock {\em Physical Review Letters}, 83(12):2402--2405, September 1999.

\bibitem{Persson2000}
B~N~J Persson and A~I Volokitin.
\newblock {Comment on “Brownian Motion of Microscopic Solids under the Action
  of Fluctuating Electromagnetic Fields”}.
\newblock {\em Physical Review Letters}, 2402(1999):3504--3504, 2000.

\bibitem{Volokitin2005}
A~I Volokitin and B~N~J Persson.
\newblock {Adsorbate-induced enhancement of electrostatic noncontact friction}.
\newblock {\em Physical Review Letters}, 94(8):86104, March 2005.

\bibitem{Kuehn2006}
Seppe Kuehn, Roger Loring, and John Marohn.
\newblock {Dielectric Fluctuations and the Origins of Noncontact Friction}.
\newblock {\em Physical Review Letters}, 96(15):1--4, April 2006.

\bibitem{Jones2003}
M.~Jones, C.~Vale, D.~Sahagun, B.~Hall, and E.~Hinds.
\newblock {Spin Coupling between Cold Atoms and the Thermal Fluctuations of a
  Metal Surface}.
\newblock {\em Physical Review Letters}, 91(8):1--4, August 2003.

\bibitem{Henkel1999}
C.~Henkel, S.~P\"{o}tting, and M.~Wilkens.
\newblock {Loss and heating of particles in small and noisy traps}.
\newblock {\em Applied Physics B: Lasers and Optics}, 69(5-6):379--387,
  December 1999.

\bibitem{Wesenberg2007}
J~H Wesenberg, R~J Epstein, D~Leibfried, R~B Blakestad, J~Britton, J~P Home,
  W~M Itano, J~D Jost, E~Knill, C~Langer, and Others.
\newblock {Fluorescence during Doppler cooling of a single trapped atom}.
\newblock {\em Physical Review A}, 76(5):53416, 2007.

\bibitem{Narayanan2010}
S.~Narayanan and \textit{et al.}
\newblock {in preparaton}, 2010.

\bibitem{Leibfried2003}
D.~Leibfried, R.~Blatt, C.~Monroe, and D.~Wineland.
\newblock {Quantum dynamics of single trapped ions}.
\newblock {\em Reviews of Modern Physics}, 75(1):281–324, 2003.

\bibitem{Shaikh2010}
F.~Shaikh and A.~Ozakin.
\newblock {Ion Motion Stability in Asymmetric Surface Electrode Ion Traps.} 
\newblock {2010 DAMOP Meeting}, May 2010.

\bibitem{Lindberg1986}
Markus Lindberg and Juha Javanainen.
\newblock {Temperature of a laser-cooled trapped three-level ion}.
\newblock {\em Journal of the Optical Society of America B}, 3(7):1008, July
  1986.

\bibitem{Siemers1992}
I.~Siemers, M~Schubert, R.~Blatt, W.~Neuhauser, and P.~E. Toschek.
\newblock {The `Trapped State' of a Trapped Ion-Line Shifts and Shape .}
\newblock {\em Europhysics Letters}, 139, 1992.

\bibitem{Reiss1996}
Dirk Reiss, Albrecht Lindner, and Rainer Blatt.
\newblock {Cooling of trapped multilevel ions: A numerical analysis}.
\newblock {\em Physical Review A}, 54(6):5133--5140, December 1996.

\bibitem{Epstein2007}
R~J Epstein, S~Seidelin, D~Leibfried, J~H Wesenberg, J~J Bollinger, J~M Amini,
  R~B Blakestad, J~Britton, J~P Home, W~M Itano, J~D Jost, E~Knill, C~Langer,
  R~Ozeri, N~Shiga, and D~J Wineland.
\newblock {Simplified motional heating rate measurements of trapped ions}.
\newblock {\em Physical Review A}, 76:33411, 2007.

\bibitem{Diedrich1989}
F~Diedrich, J~C Bergquist, W~M Itano, and D~J Wineland.
\newblock {Laser cooling to the zero-point energy of motion.}
\newblock {\em Physical Review Letters}, 62:403--406, 1989.

\bibitem{Roos1999}
Ch. Roos, Th. Zeiger, H~Rohde, H~C N\"{a}gerl, J~Eschner, D~Leibfried,
  F~Schmidt-Kaler, and R~Blatt.
\newblock {Quantum state engineering on an optical transition and decoherence
  in a Paul trap}.
\newblock {\em Physical Review Letters}, 83:4713, 1999.

\bibitem{Tamm2000}
Chr Tamm, D~Engelke, and V~B\"uhner.
\newblock {Spectroscopy of the electric-quadrupole transition
  $^{2}$S$_{1/2}$(F=0)–$^{2}$D$_{3/2}$(F=2) in
  trapped $^{171}$Yb$^{+}$}.
\newblock {\em Physical Review A}, 61(5):053405, April 2000.

\bibitem{DesLauriers2004}
L~Deslauriers, P~C Haljan, P~J Lee, K.-A. Brickman, B~B Blinov, M~J Madsen, and
  C~Monroe.
\newblock {Zero-point cooling and low heating of trapped
  $^{111}$Cd$^{+}$ ions}.
\newblock {\em Physical Review A}, 70:43408, 2004.

\bibitem{DesLauriers2006a}
L~Deslauriers, S~Olmschenk, D~Stick, W~K Hensinger, J~Sterk, and C~Monroe.
\newblock {Scaling and suppression of anomalous heating in ion traps.}
\newblock {\em Physical Review Letters}, 97(10):103007, 2006.

\bibitem{Stick2006}
D~Stick, W~K Hensinger, S~Olmschenk, M~J Madsen, K~Schwab, and C~Monroe.
\newblock {Ion Trap in a Semiconductor Chip}.
\newblock {\em Nature Physics}, 2:36, 2006.

\bibitem{Lucas2007}
D~M Lucas, B~C Keitch, J~P Home, G~Imreh, M~J McDonnell, D~N Stacey, D~J Szwer,
  and A~M Steane.
\newblock {A long-lived memory qubit on a low-decoherence quantum bus}.
\newblock {\em arXiv: 0710.4421v1}, 2007.

\bibitem{Schulz2008a}
Stephan~A Schulz, Ulrich Poschinger, Frank Ziesel, and Ferdinand Schmidt-Kaler.
\newblock {Sideband cooling and coherent dynamics in a microchip
  multi-segmented ion trap}.
\newblock {\em New Journal of Physics}, 10(4):045007, April 2008.

\bibitem{Dutta1981}
P~Dutta and P~M Horn.
\newblock {Low-frequency fluctuations in solids: 1/f noise}.
\newblock {\em Reviews of Modern Physics}, 53:497--516, 1981.

\bibitem{Sommerfeld1949}
A~Sommerfeld.
\newblock {\em {Partial Differential Equations in Physics}}.
\newblock Academic Press, New York and London, 1949.

\bibitem{Feibelman1982}
P~Feibelman.
\newblock {Surface electromagnetic fields}.
\newblock {\em Progress in Surface Science}, 12(4):287--407, 1982.

\bibitem{Varma1982}
C~M Varma, R~C Dynes, and J~R Banavar.
\newblock {Thermally created tunnelling states in glasses}.
\newblock {\em Journal of Physics C: Solid State Physics}, 15(35):L1221, 1982.

\end{thebibliography}

\end{document}